# Unbundling Open Access dimensions: a conceptual discussion to reduce terminology inconsistencies[1]

Alberto Martín-Martín*, Rodrigo Costas**, Thed N. van Leeuwen*** and Emilio Delgado López-Cózar*

*albertomartin@ugr.es; edelgado@ugr.es
Facultad de Comunicación y Documentación, Universidad de Granada, Colegio Máximo de Cartuja s/n, 18071 (Spain)

** rcostas@cwts.leidenuniv.nl
CWTS, Leiden University, Wassenaarseweg 62a, Leiden, (The Netherlands) and the DST-NRF Centre of Excellence in Scientometrics and Science, Technology and Innovation Policy, Stellenbosch University (South Africa)

*** leeuwen@cwts.leidenuniv.nl
CWTS, Leiden University, Wassenaarseweg 62a, Leiden, (The Netherlands)

**Introduction**

The beginnings of the Open Access (OA) movement are usually dated to the year 2002, when the *Budapest Open Access Initiative* (BOAI) declaration (Chan et al., 2002) was published. This was the first time the term "Open Access" was used, although the practices described in that document had already been taking place in some scientific communities long before that date. Shortly afterwards, the *Bethesda Statement on Open Access publishing* (Brown et al., 2003), and the *Berlin Declaration on Open Access to Knowledge in the Sciences and Humanities* (Max-Planck-Gesellschaft, 2003) were also published. The definitions of OA that these declarations contain are sometimes called the *BBB* definition of OA (Suber, 2008b).

From the very beginnings of the movement, it was acknowledged that there could be several ways in which a document could be made OA. The BOAI declaration recommended two routes: author self-archiving (the process by which an author deposits a copy of his/her work in an open access repository), and publishing in OA journals. Interestingly, the Bethesda and Berlin declarations did not include publishing in OA journals in their definitions of OA. Publishing in a OA journal would later be known as the *Gold route* of OA publishing, and self-archiving would be known as the *Green route*, because journals "have given their authors the green light to self-archive" (Harnad et al., 2004).

In the following years, as OA publishing started to be put into practice by academic publishers and researchers, it became evident that in the real world, OA publishing was taking forms that did not completely adhere to the *BBB*. Peter Suber and Stevan Harnad, original signatories of the BOAI and the Bethesda declarations (and considered by many the leaders of the OA movement) addressed this issue (Suber, 2008a). They realised that in many cases, OA publishing was succeeding in removing the *price barriers* for accessing documents, but they were not succeeding in removing the *permission barriers* described in the *BBB* (reuse, copy, redistribution…). Therefore, they decided to differentiate between *weak* OA (that which removes *price barriers* but not *permission barriers*), and *strong* OA (that which removes *price barriers,* and at least some *permission barriers*). Shortly after, Suber declared that he had realized that the term *weak* OA "was needlessly pejorative" and therefore decided to use a different terminology, borrowed from the field of open source software (Suber, 2008b): *gratis* OA (to replace *weak* OA), and *libre* OA (to replace *strong* OA).

[1] Alberto Martín-Martín enjoys a four-year doctoral fellowship (FPU2013/05863) granted by the Ministerio de Educación, Cultura, y Deportes (Spain). Funding from the South African DST-NRF Centre of Excellence in Scientometrics and Science, Technology and Innovation Policy (SciSTIP) is also acknowledged.

Since then, OA-related terminology has expanded even further, trying to keep up with all the variants of OA publishing that are out there. The concepts of *Diamond* and *Platinum* OA journals (Fuchs & Sandoval, 2013; Haschak, 2007) were introduced to differentiate *Gold* OA journals that charge Article Processing Charges (APC) from those that do not. The term *Hybrid* journal (Prosser, 2003; Walker, 1998) became popular to describe journals that, although maintaining the traditional model of subscriptions, gave authors the choice of publishing their articles as OA immediately upon publication after payment of an APC. Other journals adopted a *Delayed* OA model, meaning that they would make their articles OA after an specified embargo period. Some journals may decide to make some or all their articles freely accessible for a period of time for *promotional* purposes. The term *Bronze* OA has been recently used to describe documents that are "free to read on the publisher page, but without an (sic) clearly identifiable license" (Piwowar et al., 2018). Lastly, the irruption of the copyright-infringing site Sci-Hub in the world of scholarly publishing has led some to talk about *Robin Hood* OA, *Rogue* OA, and *Black* OA (Archambault et al., 2014; Björk, 2017; Green, 2017).

The myriad subtypes of OA, and the inconsistent and arbitrary ways to refer to them complicate the communication about OA-related issues. For example, the *Gold* label, which was originally intended to mean only that the venue of publication was a journal (Suber, 2008b), is now bundled with other assumptions, namely, that the article is made OA immediately upon publication, and that the article has a license that provides ample rights to all users. There is not a widespread agreement, however, on whether *Gold* OA requires the payment of an APC. This depends on whether *Diamond*/*Platinum* OA journals are considered a subset of *Gold*, or a separate type of OA. In either case, it would be an arbitrary decision that is not initially self-evident.

Another common source of confusion is that there are no fundamental differences between articles published in *Gold* OA journals, and articles published as OA in *Hybrid* journals. This evidences that these terms were designed from the perspective of the journal, and not from the perspective of the document themselves.

Turning to *Green* OA, self-archiving has been, for the most part, stifled by the ever-increasing limitations imposed by publishers regarding how, where, and when self-archiving is permitted (Gadd & Troll Covey, 2016; Kingsley, 2013). Most publishers now impose embargo periods that prevent authors from self-archiving their work immediately upon initial publication. Despite the fact that this embargo to make articles accessible from repositories implies a delay, the label *Delayed* OA is usually reserved to describe journals that make their articles freely accessible from their website after an embargo period.

Lastly, the last couple of years have seen an increasing interest in preprint publishing, as well as in alternative models of peer-review to those used in traditional journals. In the year 2017 we witnessed an explosion in the growth of the so-called preprint servers, largely enabled by the infrastructure developed by the Open Science Framework[2], a project launched by the Center for Open Science, which is a non-profit organization founded in 2013 to "increase the openness, integrity, and reproducibility of scientific research". In the field of peer-review, newly created publishing platforms such as PeerJ[3] and F1000[4] are experimenting with models of open peer-review. Other platforms, such as PubPeer[5], or scienceopen.com[6] provide readers with the opportunity to engage in post-publication, open peer-review. Furthermore, funders like the Bill & Melinda Gates Foundation are implementing their own publishing platforms, which provide the opportunity to grantees to publish their research as OA and have it peer-reviewed without the need of traditional journals (Poynder, 2018). Similarly, the European Commision recently published a tender for an Open Research Publishing Platform which also will include the necessary features to publish preprints and carry out open peer-review[7].

In light of the changes that OA publishing has gone through the last few years, and the foreseeable changes through which it will go in the coming years, we believe, that, as Suber and Harnad did ten years ago (Suber, 2008a), it is time to rethink again, and consider the best ways to interact with the concept of OA and all its variants. We believe that the original *Gold/Green* and *Libre/Gratis* distinctions are no longer capable of capturing all the permutations and nuances of OA that exist nowadays. We are not alone in this belief (Danowski, 2018; Neylon, 2013). Therefore, we think it is necessary to unbundle all the different aspects of OA and consider them as separate dimensions. We believe this exercise in abstraction will provide a tool that stakeholders such as funding bodies and policy makers can use in their discussions, and when they need to decide exactly what the minimum requirements (or the ideal requirements) for OA should be in each specific scenario.

The model we propose features six dimensions that, in combination, we believe are able to provide all the different configurations of OA that currently exist. These dimension are rooted on the concepts of legality and sustainability already expressed in van Leeuwen et al (2017) and in Martín-Martín et al. (2018), and thus all OA configurations should have some legal basis (i.e. excluding illegal forms of OA) as well as some form of sustainability (i.e. some stability and support over time). These dimensions are: prestige, user rights, stability, immediacy, peer-review, and cost (Figure 1). Each dimension allows for a range of different options depending on its nature. Furthermore, these options can be sorted by their desirability. In Figure 1, the options that are best aligned with the principles of OA development are found closer to the center of the circumference, while the options that do not align as well are placed closer to the outer rim of the circumference. It should be borne in mind that Figure 1 only displays only two opposing options for each of the dimensions, but in practice there are a whole range of options for each dimension. Therefore, the options within a dimension should not be considered to be binary. Each dimension will now be discussed.

---

[2] https://osf.io/
[3] https://peerj.com
[4] https://f1000research.com
[5] https://pubpeer.com/
[6] https://www.scienceopen.com/
[7] https://etendering.ted.europa.eu/cft/cft-document.html?docId=37014

Figure 1. Six dimensions of Open Access

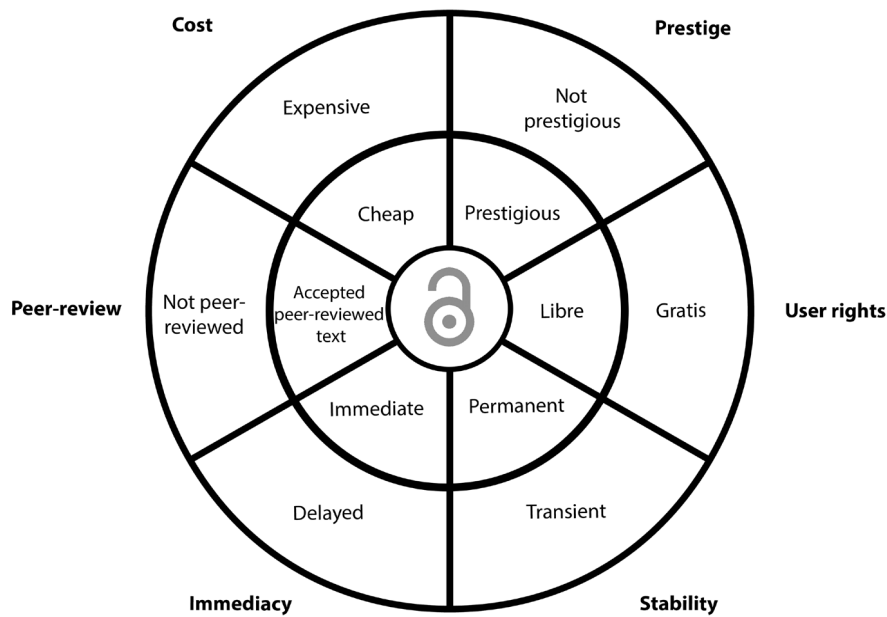

**Prestige**
This dimension is the most similar to the traditional *Gold*/*Green* distinction because it has to do with the venue where the document is published. When the *BBB* were published, and still today, the journal model is the one that prevails. In each scientific field there are some journals that through time have attained a level of prestige in their respective fields. The reason for this is that for a long time, journals were one of the few places to which scientific documents could submitted that ensured a proper vetting of their contents, and a reasonable level of dissemination among the researchers of the same field. On the other hand, the recently created web repositories are not regarded with the same level of prestige. For this reason, the *Gold*/*Green* distinction was a reasonable one.

However, reality is changing. Journals are no longer the only venue in which science can be vetted, published, and disseminated. If we unbundle these aspects from the current journal model, what is left is the prestige of the venue. In the future, some non-journal venues (for example the publishing platforms that some funding bodies and governments are currently developing) could also be considered authoritative.

**User rights**
This dimension is the one that Suber described as *Libre*/*Gratis* (Suber, 2008b). It has to do with the use license that is attached to the OA document. As Suber describes, *Gratis* OA removes *price barriers* to access the document, but does not remove *permission barriers*. *Libre* OA removes *price barriers* and at least some *permission barriers*. Therefore, there are degrees of *Libre* OA, depending on how many of the *permission barriers* are removed. Some of the *permission barriers* are: downloading the document, making copies of the document, redistributing the document, making derivative works based on the document, using the document for data mining. Creative Commons[8] licenses are one of the most popular licenses for OA documents, and enable authors to choose between more permissible user licenses and more strict ones. User rights (such as data mining) can be facilitated to a higher or lower degree by technological means. For these reason, an important aspect which should also be considered within this dimension is the level of technological affordances offered alongside the paper itself.

---

[8] https://creativecommons.org/

**Stability**
This dimension is concerned with the technical aspects of depositing a document in the internet and ensuring it remains accessible in perpetuity (it is strongly related to the idea of sustainability of the access). Nowadays there are many options to make a document freely accessible on the Web. However, some of them present more guarantees than others as regards the long-term preservation of the documents.

Well-established journals and repositories, for example, have contingency plans to preserve accessibility to their documents should they have to shut down unexpectedly, or in the event of a catastrophe. These contingency plans can take many forms. Some participate in programs such as LOCKSS[9] (Lots of Copies Keep Stuff Safe), or CLOCKSS[10] (Controlled LOCKSS). Others, like the Open Science Framework, have declared they have set aside a preservation fund which ensures that, even if the project had to close down, the data that is stored in their servers would be accessible for 50 years[11].

Stability also has to do with pledging to maintain the academic record. Journals and repositories such as ArXiv or SocArxiv maintain the academic record. This means that once a document has been published, it cannot be unpublished, by the authors or anyone else. If the authors or the journal believes that for some reason a published document should no longer be used, they can issue a retraction notice, but the original document will still be accessible. This pledge sets journals and repositories apart from other platforms that can provide access to scientific documents, such as academic social networks (ResearchGate, Academia.edu), or private websites managed by the authors (Martín-Martín, Costas, van Leeuwen, & Delgado López-Cózar, 2018). In these platforms there is no guarantee that the academic record is going to be maintained, because authors retain the power to remove any document they have uploaded. In the case of ResearchGate, the platform itself removes the full texts of documents uploaded by users who decide to delete their account from the service.

In order for a OA document to be considered *permanent* the venue of publication should have some kind of contingency plan in place, pledge to maintain the academic record, and last but not least, declare their commitment to provide continued free access to the document in question. *Permanent* documents, therefore, are those for which there exist reasonable guarantees that they will remain freely accessible on the long term. If the documents do not meet one or more of these criteria, there is no guarantee that they will be freely accessible on the long term, and therefore they should be considered *transient*. For example, several studies have found that a large portion of the documents that are made freely accessible by journal publishers do not declare any kind of OA-compatible use license (Martín-Martín et al., 2018; Piwowar et al., 2018). This is a precarious situation, because even if publishers' original intention is to maintain free access status in perpetuity, as sole copyright holders nothing could stop them if they decided to revoke that status in the future.

---

[9] https://www.lockss.org/
[10] https://www.clockss.org
[11] https://cos.io/our-products/osf/

**Immediacy**
The dimension of immediacy has to do with the amount of time that passes between the initial publication of a document, and the moment it is made freely accessible. At first glance, this dimension seems to be straightforward: the sooner the document is made freely accessible, the more *immediate* its OA status becomes. Likewise, the longer it takes a document to be made freely accessible, whether because of embargo policies, or any other reasons, the more *delayed* its OA status will be. However, it is worth noting that it is not entirely clear what "initial publication" nowadays means: for journal articles for example, is it the moment of acceptance? Is it the moment the document is first made available online (freely accessible or not)? Or is it the moment the article is included in a volume and issue? (Haustein, Bowman, & Costas, 2015)

**Peer-review**
Providing a comprehensive model of all possible peer-review implementations (i.e. pre-publication or post-publication, closed or open, editorially managed or accidental, etc.) is not within the scope of this proposal. Nevertheless, peer-review remains a key issue that needs to be considered in OA publishing. In some cases, funding bodies or policy makers might consider that peer-review is not necessary to fulfil their publication requirements. The current EU tender for an Open Research Publishing Platform, for example, considers the possibility of publishing preprints in this future platform without the intention to carry out peer-review and final publication in the same platform. The European Commission will continue encouraging early sharing of preprints in the upcoming *Horizon Europe* programme (European Commission, 2018), although it is still unknown whether preprint sharing will satisfy their OA requirement.

**Cost**
Last but not least, the cost factor should not be overlooked. In this respect, the current OA publishing system is dominated by large publishers, which have found in APCs an additional stream of revenue. In this new scenario, each country has reacted in its own way. In 2012, the United Kingdom decided to follow the *Gold* route of OA, which today we know has had as a consequence an steady increase of 11% per annum from 2013 to 2016 in the cost of journal subscriptions and APC expenditure (Tickell et al., 2017). Other countries are currently locked in arguments with large publishers over the costs of renewing their subscription and publishing agreements (Kwon, 2018; Vogel, 2017). The Indian government, for its part, decided to "exclude from reviews for promotion any articles published in journals that charge a processing fee" (Padma, 2017), in an attempt to eradicate the problem of fraudulent scientific journals. The European Commission is planning to stop covering APCs for articles published in Hybrid journals (which are often more expensive than those of full OA journals) in the forthcoming *Horizon Europe* programme (European Commission, 2018).

**Conclusions**
This study proposes to move away from arbitrary and inconsistent OA terminology (such as colour or metal-based labels, or ambiguous terms such as *Hybrid*) by identifying the main dimensions that affect OA publishing. In the presentation of this proposal at the STI conference, we would further elaborate on how this model maps to current OA terminology.